\documentclass[utf8]{frontiersinFPHY_FAMS} 

\setcitestyle{square} 
\usepackage{url,hyperref,microtype,subcaption}
\usepackage[onehalfspacing]{setspace}
\usepackage{bm}

\def\keyFont{\fontsize{8}{11}\helveticabold }
\def\firstAuthorLast{Hayen} 
\def\Authors{Leendert Hayen\,$^{*}$}
\def\Address{Universit\'e de Caen Normandie, ENSICAEN, CNRS/IN2P3, LPC Caen, UMR6534, 14000 Caen, France}
\def\corrAuthor{Leendert Hayen}

\def\corrEmail{hayen@lpccaen.in2p3.fr}

\begin{document}
\firstpage{1}

\title[Form factors in precision $\beta$ decay]{The form factor expansion in the precision $\beta$ decay era} 

\author[\firstAuthorLast ]{\Authors} 
\address{} 
\correspondance{} 

\extraAuth{}

\maketitle

\begin{abstract}

\section{}

Precision tests of the Standard Model using $\beta$ decay have always relied on a careful choice of transition to minimize residual nuclear structure uncertainties. Following breakthroughs in nucleon-level radiative corrections in the last decade, however, corrections due to nuclear structure are once more a limiting factor in several scenarios. Progress in \textit{ab initio} nuclear theory provides a path forward, but common recoil-order approximations in traditional formalisms often go unnoticed. Here, we critically examine their origin and address recently resolved and identify open questions.

\tiny
 \keyFont{ \section{Keywords:} nuclear $\beta$ decay, multipole expansion, recoil corrections, radiative corrections, Standard Model test} 
\end{abstract}

\section{Introduction}
\label{sec:introduction}

Much like the cyclicity of history, so is the study of nuclear structure corrections in precision beta decays once more a fashionable topic. Following intensive efforts between the early 1950's to late 1970's \cite{Biedenharn1953, Rose1954, Wu1964, Fujii1959, Morita1959, Morita1958, Morita1963, Stech1964, Schulke1964, Weidenmuller1961, Kim1965, Kim1965a, Armstrong1972a, Behrens1971, Konopinski1966, Holstein1974, Behrens1982, Morita1985}, theory efforts on formalism development were sufficiently advanced to outstrip experimental progress for the following decades \cite{Severijns2006, Herczeg2001, Cirigliano2013b, Severijns2011}. Recently, however, the advent of nuclear \textit{ab initio} methods have reinvigorated detailed treatments of the nuclear weak currents \cite{Ekstrom2014, Sargsyan2026, Gysbers2019, Acharya2025}, and in the process of 'rediscovery' translate it into methods suited for modern effective field theories \cite{Hammer2019, Epelbaum2020}. Simultaneously, the advent of novel detection schemes and experimental techniques - including atom/ion traps \cite{Behr2005, Behr2008} and, more recently, cyclotron radiation emission spectroscopy \cite{Oblath2024, Byron2023} and a variety of `quantum sensors' \cite{Carney2023, Smolsky2024, Friedrich2021} - have reinforced the need for precision predictions \cite{Hayen2018, Hayen2019a, Hayen2024}. Finally, significant advances in neutrino detection have underlined the necessity of precision beta spectrum predictions in the reactor antineutrino anomaly \cite{Hayen2019, Hayen2019b, Zhang2025} and, looking forward, (coherent) neutrino scattering \cite{Hayen2025}.

The complexity resides in a systematic treatment of the nuclear electroweak response, and various authors have used a number of different methods to break it down into manageable pieces. While each carries specific strengths, so does each comes with a number of approximations. As some of these formalisms are picked back up by the modern theory community, the generational gap in expertise means some of the aforementioned approximations have not received as much attention as would be desirable. In this work, we will go over the basic philosophies, and focus on the multipole decomposition and its (oft-neglected) approximations. We will give recent examples of careful resolution of double-counting and other sorts of errors, and reflect upon potential issues.

\section{Decomposition strategy}
\label{sec:decomposition_strategy}

For momentum transfers $q$ much smaller than the $W$ boson mass, the $T$-matrix element for $\beta$ decay reduces to a simple contact interaction
\begin{align}
    T = \frac{G_F}{\sqrt{2}}L_\mu\int &d\bm{r}~  e^{-i\bm{q}\cdot\bm{r}}\langle f| V^\mu-A^\mu|i\rangle + h.c.,
    \label{eq:T_matrix}
\end{align}
where $L_\mu = \bar{u}_e\gamma_\mu(1-\gamma^5)v_\nu$, $l_\mu(\bm{r}) = e^{-i\bm{q}\cdot\bm{r}}L_\mu$ from translation invariance with $\bm{q} = \bm{p}_e+\bm{p}_\nu$, and $V^\mu$ and $A^\mu$ are the general vector and axial vector currents, respectively, acting on initial and final nuclear wave functions.


\subsection{Form factor decomposition}

The main question is how to describe the nuclear electroweak response by a covariant combination of form factors. Initial efforts on a description of $\beta$ decay \cite{Konopinski1935, Konopinski1941, Konopinski1943} revolved mostly around an expansion of the lepton plane wave functions, $e^{-i\bm{q}\cdot\bm{r}}$, and identifying leading-order nuclear matrix elements. Later, two main philosophies emerged. The first is the `elementary particle' treatment, which takes inspiration from the on-shell spin-1/2 covariant decomposition,
\begin{subequations}
\begin{align}
    &\langle p^\prime|V^\mu-A^\mu|p\rangle = \bar{u}(p^\prime) [f_V\gamma^\mu-i\frac{f_M}{2M}\sigma^{\mu\nu}q_\nu-f_Sq^\mu \nonumber \\
    &+g_A\gamma^\mu\gamma^5-i\frac{g_T}{2M}\sigma^{\mu\nu}q_\nu\gamma^5-i\frac{g_P}{2M}\gamma^5q^\mu{\Large]}u(p)
    \label{eq:nucleon_weak_current}
\end{align}
\end{subequations}
where all form factors are dimensionless functions of $q^2=(p^\prime-p)^2$. For simple $\beta$ decays a covariant version of the $T$ matrix element of Eq. \ref{eq:T_matrix}) was first introduced by Kim and Primakoff \cite{Kim1965, Kim1965a}, and later expanded upon by Holstein \cite{Holstein1974}. It has the distinct benefit that all its expressions are manifestly Lorentz-invariant and its form factors obey simple symmetry relations, lending itself easily to evaluation in different frames (particularly useful for, e.g., beta-delayed particle emission) and greatly simplified expressions for decays between highly symmetric states such as superallowed $0^+$ to $0^+$ decays or mirror (isospin $T=1/2$) decays. An extension to forbidden decays is particularly cumbersome, however, due to the large number of terms appearing and little effort has gone into this direction \cite{Kim1965}. Nonetheless, many experimentalists are well-acquainted with the Holstein formalism for the aforementioned reasons and to this day many analyses use their expressions, as discussed in Sec. \ref{sec:double_counting_Vud} \footnote{As an aside, the `Coulombic' corrections due to soft photon exchange were developed only to first order \cite{Holstein1974a, Holstein1974b, Calaprice1976} meaning their use is limited mostly to low-$Z$ nuclei.} \cite{Kleppinger1977, Gorelov2005, Wauters2010, Soti2014, Brown2018, Burkey2022, Severijns2023, Sargsyan2022}.

The second philosophy has been to perform a relativistic multipole decomposition similar to what is standard in non-relativistic electromagnetism. The question as to whether such an expansion exists had been answered by Durand et al. \cite{Durand1962}, and various others \cite{Galindo1969, Delorme1979}. Crucially, these authors showed that a \textit{clean decomposition exists only in the Breit frame} using the helicity formalism developed by Jacob and Wick \cite{Jacob1959}. This is a critical point, and we will therefore dwell on it a little.

The central idea of a multipole expansion is the construction of irreducible tensors that respect the conservation of angular momentum of the nuclear states in their respective rest frames. This can easily be seen from Eq. (\ref{eq:T_matrix}) using the spherical harmonic expansion of the exponential,
\begin{equation}
    e^{-i\bm{q}\cdot\bm{r}} = 4\pi \sum_{L,M} (-1)^{L}j_L(qr)Y_{LM}(\hat{r})Y^*_{LM}(\hat{q}),
    \label{eq:exp_spherical}
\end{equation}
with $j_L$ a spherical Bessel function. While Eq. (\ref{eq:exp_spherical}) is valid for any two three-vectors, the $\bm{r}$ we are interested in are the internal nuclear coordinates as the center-of-mass motion results simply in an overall phase. As a consequence, for the decomposition into irreducible spherical tensors to be valid, $\bm{q}$ must also be an \textit{internal} momentum transfer rather than, e.g., the center-of-mass motion.

In the lab frame, we have $q^{l} = (E_f-M_i, \bm{p}_f^{l})$. The somewhat unphysical Breit frame, chosen as the frame where $\bm{p}_i+\bm{p}_f=0$, has $q^{B}=(M_f-M_i, \bm{q}^{B})$ with $\bm{q}^{B} = 2\bm{p}_f^{B}$. Crucially, $\bm{q}^B$ corresponds to \textit{internal} momentum transfer, and both initial and final states can be described in the same frame by a rotation. The approach by Donnelly, Walecka, \textit{et al.} \cite{DeForest1966, Donnelly1975, Walecka2004} (DW) proceeds simply by writing the nuclear current as $V^\mu-A^\mu \equiv J^\mu = (\rho, \bm{J})$, and writing $\bm{J}^B$ in spherical components, $\bm{J}^B = \sum_\lambda J^B(\lambda)\bm{e}_\lambda^*$ ($\lambda=0,\pm1$). Taking $\bm{q}$ to lie along the $z$-axis and combining Eq. (\ref{eq:exp_spherical}) results in the usual multipoles (Coulomb, Longitudinal, Tranverse Electric and Tranverse Magnetic) after some algebraic manipulation \cite{Walecka2004}. This formalism is popular among nuclear \textit{ab initio} theorists \cite{King2022, Glick-Magid2022} and expressions for the electroweak currents can be plugged in directly from, e.g., a chiral EFT approach (see, e.g., Appendix E in \cite{King2022}).

A separate part of the community - including the work by Towner and Hardy \cite{Hardy2020} used for $V_{ud}$ extraction of superallowed decays; see Sec. \ref{sec:double_counting_Vud} - uses a multipole expansion first written by Stech and Sch\"ulke \cite{Stech1964, Schulke1964} and expanded upon by Behrens and B\"uhring (BB) \cite{Behrens1971, Behrens1982}. It makes explicit mention of the Breit frame reduction and instead uses a spherical tensor decomposition of $J^B_\mu(0)$ as follows
\begin{align}
    \langle f | \rho^B|i\rangle &= \sum_{L,M}(-)^{J_f-M_f}\left(\begin{array}{ccc}
        J_f &  L & J_i \\
        -M_F & M & M_i 
    \end{array}\right) \nonumber \\
    &\times \hat{J_i}Y_{LM}^*(\hat{q})F_L(q^2) \\
    \langle f | \bm{J}^B|i\rangle &= \sum_{K,L,M}(-)^{J_f-M_f}\left(\begin{array}{ccc}
        J_f &  K & J_i \\
        -M_F & M & M_i 
    \end{array}\right) \nonumber \\
    &\times \hat{J_i}\bm{Y}_{KLM}^*(\hat{q})F_{KL}(q^2)
\end{align}
with $\hat{j}=\sqrt{2j+1}$, the symbol in parentheses is a Wigner-3j symbol, $Y^M_L$ ($\bm{Y}^M_{KL}$) are (vector) spherical harmonics \cite{Edmonds1957}, and $j_L$ is absorbed into the form factor. Keeping the Wigner-Eckart theorem in mind, it is clear that the form factors can be thought of as reduced matrix elements. The BB and DW approaches are clearly related, and explicit translation tables are given in Refs. \cite{Behrens1982, Hayen2020a, King2022}. All experimental observables can be written in terms of the form factors and their explicit angular momentum transformation facilitates final calculations using simple group theory methods. Unlike the covariant description of Holstein, on the other hand, angular correlations in $\beta$-delayed particle emission results in some tedious bookkeeping for frame-to-frame transformations. 

\subsection{Non-relativistic operator reduction and operator subtleties}

While experiments can be used to obtain form factors independently, in order to provide theoretical predictions one reduces them to nuclear matrix elements. Historically, this has been done using the impulse approximation which implies that the total nuclear current is taken to be a sum of individual nucleon currents according to Eq. (\ref{eq:nucleon_weak_current}). Current \textit{ab initio} methods, on the other hand, can directly construct the electroweak response to a given chiral order and naturally include many-body currents.

Given that almost all calculations currently entering the nuclear extraction of $V_{ud}$ (as well as many angular correlation measurements) use the historic approach, we outline subtle differences between methods. All methods perform some kind of non-relativistic reduction of the operators of Eq. (\ref{eq:nucleon_weak_current}) to act on non-relativistic nucleon wave functions. In the Behrens-B\"uhring formalism, one considers individual nucleons inside some mean field potential, $V(r)$, and keeps a relativistic spinor notation for individual nucleons as an intermediate step. In the non-relativistic limit of the Dirac equation, one can write the small component as a function of the large component in a Schr\"odinger-like equation which depends explicitly on this mean field potential \cite{Behrens1982}. In this formalism, therefore, off-diagonal operators (i.e. those connecting large and small radial components of a Dirac spinor) result in an explicit dependence on the nuclear mean-field potential \cite{Behrens1978}. This is somewhat counter-intuitive to the impulse approximation (and doesn't appear in chiral-EFT inspired approaches) but generally obtains very good agreement with data\footnote{This approach is typically used with phenomenological many-body methods such as the nuclear shell model, which are fitted to various experimental observables. As such, it folds in some effects that would show up as, e.g., two-body currents in a $\chi$EFT approach.}. Importantly, the operators must be evaluated in the Breit frame to be consistent with the multipole expansion. Other approaches, such as that by Holstein \cite{Holstein1974} and Rose et al. \cite{Rose1954, Rose1954a}, use a different systematic expansion. In 1950, Foldy and Wouthuysen (FW) \cite{Foldy1950} proposed a canonical transformation to, order-by-order in $1/m$, put a Hamiltonian in block-diagonal form acting on Pauli spinor wave functions with eigenvalues $\pm(E^2+p^2)^{1/2}$. Using the same transformation on the $\beta$ decay operators, one obtains the order-by-order corrected operator that may then be used in the form factor expansion. The two approaches give results in general agreement but differ subtly on a conceptual level. The Behrens-B\'uhring approach considers free \textit{quasi}particles, in the sense that they are independent particles in a mean-field potential. The FW transformation, on the other hand, considers \textit{actually free} particles, as the canonical transformation starts from the free Dirac Hamiltonian.

\subsection{Multipole decomposition and recoil corrections}

As stated \textit{ad nauseam} above, the multipole decomposition is valid only in the \textit{Breit} frame. The lepton current must therefore either be evaluated in the same frame, or some Lorentz transformation must be included. In practice, the lepton current is most often evaluated in the rest frame of the \textit{final} state with the electron/positron wave function taken as the solution to the Dirac equation with a spherical electrostatic potential\footnote{In the traditional Fermi function description the final state is also assumed infinitely massive, and additional corrections are included \textit{a posteriori} to account for the recoiling final state, see e.g. \cite{Wilkinson1982}.}. As such, one must perform a Lorentz transformation of the nuclear form factors into the rest frame of the final state. One natural consequence is the Coulombic ($\rho^B$) and Longitudinal ($J_0^B$) form factors mix with each getting a contribution from the other at order $q/M$. Despite its simplicity, it is rarely explicitly taken into account, with kinematic recoil-order corrections usually taken from, e.g., Shekhter \cite{Shekhter1959}. Besides the lack of transparency, the latter is valid only for allowed transitions and depends on the type of decay (i.e. Fermi or Gamow-Teller). Following the growing interest in (unique) forbidden decays \cite{Glick-Magid2016a, Glick-Magid2023, Hayen2019b, Seng2025}, a coherent inclusion of kinematic recoil-order corrections in modern multipole formalisms should be strongly considered.

\section{(Un)resolved issues - radiative corrections}

From the preceding sections is it clear that recoil corrections appear at various levels and proliferate through various observables in sometimes subtle ways. This has been exacerbated by the combination of results in various formalisms, which - even though it takes advantage of particular strengths of each - have made their correct application less transparent and error-prone. Quasi all current approaches using the multipole decomposition equate the lab frame with the Breit frame (implicitly or explicitly) and combine this with recoil-order corrections derived in a variety of different ways. In this section, we will go over several examples where errors have been caught only recently, sometimes with significant impact on Beyond Standard Model searches. Additionally, we will comment on some open questions that have, to the authors' knowledge, never been raised by the community and can result in potential flaws in the \textit{ab initio} radiative corrections.

\subsection{Double counting in mirror $V_{ud}$ extraction}
\label{sec:double_counting_Vud}
The first example presumably arose as a consequence of the aforementioned lack of transparency. In short, a double-counting mistake was introduced because the extraction of the Gamow-Teller/Fermi mixing ratio in mirror decays, $\rho$, from experimental data is traditionally performed in the Holstein formalism \cite{Severijns2008, Naviliat-Cuncic2009, Fenker2018, Kleppinger1977, Severijns2023, Sternberg2015, Gallant2023, Triambak2017}, and was afterwards used as an input for calculations using the Behrens-B\"uhring formalism by Towner and Hardy \cite{Hardy2004, Hardy2020}. While in many ways these give rise to identical results for allowed transitions with appropriate care, this is not the case for certain Coulomb-recoil type corrections (i.e. $\mathcal{O}\{\alpha Z F(q^2) / MR \}$ ), discussed in more detail below. In the Holstein formalism, some of these corrections were not introduced until several years later \cite{Holstein1979, Holstein1979b} and do not typically feature in experimental analyses\footnote{Note that the aforementioned experimental analyses all refer to the original 1974 work \cite{Holstein1974}.}. In the BB formalism, on the other hand, these are included by default but come out of developing the various power expansions introduced by the authors (i.e. the origin of different terms is not obvious upon first glance). Depending on the way the original analysis was performed then, an extraction of an effective $\rho_\mathrm{eff}$ is reported, and its inclusion with the BB-developed calculations contains significant double counting. This was noted only a few years ago \cite{Hayen2021}, which resulted in a shift of the extracted mirror $V_{ud}$ value by 3 standard deviations and resolved a long-standing question as to its disagreement with neutron and superallowed extractions as well as its internal consistency. Because of the explicit dependence on the precise way each experiment deals with their application of theory corrections, this is a rather insidious problem and calls for a clear and unified prescription within the community.

A similar question now arises with recent findings of additional energy dependence in the $\beta$ energy spectrum due to nuclear structure effects \cite{Gorchtein2018, Seng2019, Seng2021c, Seng2023}. As such, these contribute to the total half-life, and currently constitute the largest uncertainty in the superallowed $V_{ud}$ determination \cite{Hardy2020}. It is likely, however, that similar corrections appear in angular correlations but are currently unaccounted for (such as in mirror decays). This situation is then analogous to that of the previous paragraph, but its effects are yet to be investigated.

\subsection{Coulomb-recoil rescaling and cancellations in full $\mathcal{O}(\alpha)$ calculations}

As mentioned in the previous section, the appearance of Coulomb-recoil induced terms was only discussed explicitly in the elementary particle approach by Bottino, Ciocchetti and Kim \cite{Bottino1973b, Bottino1974}, and later by Holstein \cite{Holstein1979, Holstein1979b}, despite their presence in the multipole works by Stech and Sch\"ulke \cite{Stech1964, Schulke1964, Behrens1971, Behrens1982}. Among other corrections, it results in a `renormalization' of several form factors and can be at the level of $Z\times 0.1\%$, i.e. not negligible with respect to current precision. Coincidentally, corrections to the vector coupling are proportional to the induced scalar form factor ($f_S$ in Eq. (\ref{eq:nucleon_weak_current})), which is forced to zero through the conservation of the vector current. The axial Gamow-Teller form factor, on the other hand, receives corrections due to, among others, weak magnetism which results in the aforementioned estimate. This was noted in the context of neutron $\beta$ decay \cite{Hayen2019c} but remained unpublished and was put aside when more sophisticated treatments of the radiative corrections arrived a few years later \cite{Hayen2021, Gorchtein2021}. The latter works showed that weak magnetism, indeed, does play a significant role in the inner radiative correction of $g_A$, but is at least a factor three smaller than the traditional estimate. This is somewhat surprising, given that the weak magnetism contribution to the $\gamma W$ box similarly comes from the Born response, i.e. the same low-momentum photon exchange that results in the Coulomb interaction. This difference has not been studied explicitly, and potentially calls into question aspects of the electromagnetic corrections in the traditional multipole study.

Related but contrary to the previous point, we note that the full $\mathcal{O}(\alpha)$ calculation can result in a cancellation of some multipole contributions. This was mentioned in Ref. \cite{Hayen2021} for the induced tensor contribution to the axial Gamow-Teller coupling. Opposite contributions come from parts of the $\gamma W$ box function and the weak vertex correction. In current \textit{ab initio} treatments of radiative corrections the focus lies squarely on the evaluation of the $\gamma W$ box \cite{Seng2023}, however, and care must be taken when extending the calculation away from superallowed $0^+$ to $0^+$ transitions.

\subsection{Crossover into the quasielastic regime}

Following the groundbreaking results in 2018 for the neutron \cite{Seng2018, Gorchtein2018, Gorchtein2023}, the dispersion approach to radiative corrections elucidated the importance of the quasielastic regime in the nuclear $\gamma W$ box diagram \cite{Seng2019b}. This was contrary to the earlier work of Towner and Hardy \cite{Towner1992a, Towner1994, Towner2002}, which focused on the modification of the axial strength in low-lying nuclear states. To this day, however, a simple Fermi gas model remains the best estimate across the full data set and its uncertainty is the dominant one in the current $V_{ud}$ determination \cite{Hardy2020}. 

Recently, a first evaluation of the nuclear structure correction was evaluated for $^{10}$C using the No Core Shell Model (NCSM) \cite{Gennari2025}. These results represent a major step forward in the fidelity of the nuclear $V_{ud}$ determination, and contained a comprehensive breakdown of the uncertainty budget. While in their results nuclear shadowing effects \cite{Gorchtein2024} give rise to the dominant uncertainty, it appears that the treatment of quasielastic scattering is not completely transparent. As stated by the authors, NCSM provides bound states as well as a discretized version of the continuum. The method is not well-suited, however, to describe quasielastic processes as evidenced by significant differences in, e.g., electron scattering \cite{Lovato2023}. While they obtain a result consistent with the simple estimate, the effect of the quasielastic regime warrants more attention using more appropriate methods \cite{Rocco2019}. Quantum Monte Carlo and Green's Function Monte Carlo methods, for example, are well-suited but are computationally significantly more complex \cite{Rocco2020} and have difficulties reaching all but the lowest-mass nuclei. Methods such as the Lorentz Integral Transform \cite{Sobczyk2024} or short-time approximation \cite{King2024} can provide useful paths forward but have not yet been utilized in this context.

\subsection{Validity of recoil-free radiative corrections}

Finally, we consider recoil-order corrections in radiative corrections due the difference in momentum of the initial and final states and show up at $\mathcal{O}(\alpha Z q / M)$. Typically, these are dismissed as they are of order $\mathcal{O}(10^{-6} \times q ~\mathrm{MeV}^{-1})$. As a consequence, no attention is paid to the frame in which the multipole expansion is performed, such in the recent $^{10}$C work \cite{Gennari2025}. Based on the foregoing discussion, however, it is not clear that this approximation is valid. The multipole responses in Ref. \cite{Gennari2025} are calculated explicitly until $q \approx 500$ MeV, while true quasielastic processes can probe even higher momenta. Clearly, a naive application of the previous estimate results in unrealistic per-mille level effects. A significant part of the strength is dominated by the low- to medium-$q$ range, however, and effects will likely be (significantly) lower. 

Even so, equating the rest and Breit frames in the loop calculation clearly poses conceptual issues, exacerbated by the absence of a well-defined Breit frame for a virtual loop momentum $\bm{q}$. Another recent work \cite{Seng2025} discusses a somewhat related issue in the context of the Wigner-Eckart theorem whereby the external electron momentum is used instead as a spin-quantization axis. This is briefly discussed in classic textbooks \cite{Walecka2004}, but lies at the heart of the helicity description that was introduced in the late 1950's \cite{Jacob1959, Durand1962} and was touched upon earlier in this work. Clearly, more work is needed to elucidate this problem and significant corrections may show up in the absence of a deeper-lying cancellation.

\section{Conclusion}

In the last years years the description of nuclear structure effects in the best-understood nuclear $\beta$ decays have become a precision bottleneck, which requires a careful evaluation of the formalisms currently in use. We have outlined the main philosophies and connected it to historically used approaches. So-called recoil-order corrections enter naturally at the per-mille scale but with corrections contributing up to large momenta, a detailed treatment has become paramount. We have discussed a number of corrections that have surfaced only recently, and identified several open questions as well as potential avenues for their resolution.

As the field benefits of progress in nuclear \textit{ab initio} techniques, a careful streamlining of the current descriptions is therefore desirable. The community is slowly moving in this direction, with a new generation of theorists re-discovering and expanding upon work done well over half a century prior. It is our hope that with these improvements, nuclear structure uncertainties will no longer be a dominant uncertainty in low-energy Standard Model tests and instead fully enable and support the precision gains anticipated in novel experimental efforts.

\section*{Funding}
I acknowledge the support of the French National Agency for Research (Agence Nationale de la Recherche). 

\section*{Acknowledgments}
I would like to thank the International Union of Pure and Applied Physics (IUPAP) C12 Commission for this opportunity. I am honored to be the recipient of the 2025 IUPAP Early Career Scientist Prize in Nuclear Physics (C12). Further, I would like to acknowledge the many colleagues that I have had the pleasure to interact with in both experimental and theoretical communities, many of whom I am happy to call friends. Particular mention goes to Nathal Severijns, Albert Young, Marcella Grasso, Etienne Li\'enard and Xavier Fl\'echard as mentors that have contributed immensely to the scientist and person I am today. Finally, I would like to thank my wife and our son for their love.

\bibliographystyle{Frontiers-Vancouver} 
\bibliography{library}

\begin{thebibliography}{107}
\expandafter\ifx\csname natexlab\endcsname\relax\def\natexlab#1{#1}\fi
\expandafter\ifx\csname urlstyle\endcsname\relax
  \expandafter\ifx\csname doi\endcsname\relax
  \def\doi#1{doi:\discretionary{}{}{}#1}\fi \else
  \expandafter\ifx\csname doi\endcsname\relax
  \def\doi{doi:\discretionary{}{}{}\begingroup \urlstyle{rm}\Url}\fi \fi
\expandafter\ifx\csname selectlanguage\endcsname\relax
  \def\selectlanguage#1{}\fi

\bibitem[{Biedenharn and Rose(1953)}]{Biedenharn1953}
Biedenharn LC, Rose ME.
\newblock {Theory of Angular Correlation of Nuclear Radiations}.
\newblock {\em Reviews of Modern Physics\/} {\bf 25} (1953) 729--777.
\newblock \doi{10.1103/RevModPhys.25.729}.

\bibitem[{Rose and Osborn(1954{\natexlab{a}})}]{Rose1954}
Rose ME, Osborn RK.
\newblock {Nuclear Matrix Elements in Beta Decay}.
\newblock {\em Physical Review\/} {\bf 93} (1954{\natexlab{a}}) 1326.

\bibitem[{Wu(1964)}]{Wu1964}
Wu CS.
\newblock {The universal Fermi interaction and the conserved vector current in beta decay}.
\newblock {\em Reviews of Modern Physics\/} {\bf 36} (1964) 618--632.
\newblock \doi{10.1103/RevModPhys.36.618}.

\bibitem[{Fujii and Primakoff(1959)}]{Fujii1959}
Fujii A, Primakoff H.
\newblock Muon capture in certain light nuclei.
\newblock {\em Il Nuovo Cimento\/} {\bf 12} (1959) 327--355.
\newblock \doi{10.1007/BF02745906}.

\bibitem[{Morita(1959)}]{Morita1959}
Morita M.
\newblock {Higher Order Corrections to the Allowed Beta Decay}.
\newblock {\em Physical Review\/} {\bf 113} (1959) 1584--1589.
\newblock \doi{10.1103/PhysRev.113.1584}.

\bibitem[{Morita and Morita(1958)}]{Morita1958}
Morita M, Morita RS.
\newblock {First-Forbidden Transitions in Parity-Nonconserving Beta Decay}.
\newblock {\em Physical Review\/} {\bf 109} (1958) 2048--2058.
\newblock \doi{10.1103/PhysRev.109.2048}.

\bibitem[{Morita(1963)}]{Morita1963}
Morita M.
\newblock {Theory of Beta Decay}.
\newblock {\em Progress of Theoretical Physics Supplement\/} {\bf 26} (1963) 1--63.
\newblock \doi{10.1143/PTPS.26.1}.

\bibitem[{Stech and Sch{\"{u}}lke(1964)}]{Stech1964}
Stech B, Sch{\"{u}}lke L.
\newblock Nuclear $\beta$-decay. i.
\newblock {\em Zeitschrift f{\"{u}}r Physik\/} {\bf 179} (1964) 314--330.
\newblock \doi{10.1007/BF01381649}.

\bibitem[{Sch{\"{u}}lke(1964)}]{Schulke1964}
Sch{\"{u}}lke L.
\newblock Nuclear $\beta$-decay. ii.
\newblock {\em Zeitschrift f{\"{u}}r Physik\/} {\bf 179} (1964) 331--342.
\newblock \doi{10.1007/BF01381650}.

\bibitem[{Weidenm{\"{u}}ller(1961)}]{Weidenmuller1961}
Weidenm{\"{u}}ller HA.
\newblock {First-Forbidden Beta Decay}.
\newblock {\em Reviews of Modern Physics\/} {\bf 33} (1961) 574.

\bibitem[{Kim and Primakoff(1965{\natexlab{a}})}]{Kim1965}
Kim CW, Primakoff H.
\newblock {Application of the Goldberger-Treiman Relation to the Beta Decay of Complex Nuclei}.
\newblock {\em Physical Review\/} {\bf 139} (1965{\natexlab{a}}) B1447.

\bibitem[{Kim and Primakoff(1965{\natexlab{b}})}]{Kim1965a}
Kim CW, Primakoff H.
\newblock {Theory of Muon Capture with Initial and Final Nuclei Treated as "Elementary" Particles}.
\newblock {\em Physical Review\/} {\bf 140} (1965{\natexlab{b}}) B566--B575.
\newblock \doi{10.1103/PhysRev.140.B566}.

\bibitem[{Armstrong and Kim(1972)}]{Armstrong1972a}
Armstrong L, Kim CW.
\newblock {Comparison of Impulse-Approximation and Elementary-Particle Treatments in Nuclear Beta Decay}.
\newblock {\em Physical Review C\/} {\bf 6} (1972) 1924--1934.
\newblock \doi{10.1103/PhysRevC.6.1924}.

\bibitem[{Behrens and B{\"{u}}hring(1971)}]{Behrens1971}
Behrens H, B{\"{u}}hring W.
\newblock {Nuclear beta decay}.
\newblock {\em Nuclear Physics A\/} {\bf 162} (1971) 111--144.
\newblock \doi{10.1016/0375-9474(71)90489-1}.

\bibitem[{Konopinski(1966)}]{Konopinski1966}
Konopinski EJ.
\newblock {\em {The Theory of Beta Radioactivity}\/} (Oxford University Press) (1966).

\bibitem[{Holstein(1974{\natexlab{a}})}]{Holstein1974}
Holstein B.
\newblock {Recoil effects in allowed beta decay: the elementary particle approach}.
\newblock {\em Reviews of Modern Physics\/} {\bf 46} (1974{\natexlab{a}}) 789.

\bibitem[{Behrens and B{\"{u}}hring(1982)}]{Behrens1982}
Behrens H, B{\"{u}}hring W.
\newblock {\em {Electron radial wave functions and nuclear beta-decay}\/} (Clarendon Press, Oxford) (1982).

\bibitem[{Morita(1985)}]{Morita1985}
Morita M.
\newblock {Weak nucleon currents in beta decay and muon capture}.
\newblock {\em Hyperfine Interactions\/} {\bf 21} (1985) 143--158.
\newblock \doi{10.1007/BF02061982}.

\bibitem[{Severijns et~al.(2006)Severijns, Beck, and Naviliat-Cuncic}]{Severijns2006}
Severijns N, Beck M, Naviliat-Cuncic O.
\newblock {Tests of the standard electroweak model in nuclear beta decay}.
\newblock {\em Reviews of Modern Physics\/} {\bf 78} (2006) 991--1040.
\newblock \doi{10.1103/RevModPhys.78.991}.

\bibitem[{Herczeg(2001)}]{Herczeg2001}
Herczeg P.
\newblock {Beta decay beyond the standard model}.
\newblock {\em Progress in Particle and Nuclear Physics\/} {\bf 46} (2001) 413--457.
\newblock \doi{10.1016/S0146-6410(01)00149-1}.

\bibitem[{Cirigliano and Ramsey-Musolf(2013)}]{Cirigliano2013b}
Cirigliano V, Ramsey-Musolf MJ.
\newblock {Low energy probes of physics beyond the standard model}.
\newblock {\em Progress in Particle and Nuclear Physics\/} {\bf 71} (2013) 2--20.
\newblock \doi{10.1016/j.ppnp.2013.03.002}.

\bibitem[{Severijns and Naviliat-Cuncic(2011)}]{Severijns2011}
Severijns N, Naviliat-Cuncic O.
\newblock {Symmetry Tests in Nuclear Beta Decay}.
\newblock {\em Annual Review of Nuclear and Particle Science\/} {\bf 61} (2011) 23--46.
\newblock \doi{10.1146/annurev-nucl-102010-130410}.

\bibitem[{Ekstr{\"{o}}m et~al.(2014)Ekstr{\"{o}}m, Jansen, Wendt, Hagen, Papenbrock, Bacca et~al.}]{Ekstrom2014}
Ekstr{\"{o}}m A, Jansen GR, Wendt KA, Hagen G, Papenbrock T, Bacca S, et~al.
\newblock {Effects of Three-Nucleon Forces and Two-Body Currents on Gamow-Teller Strengths}.
\newblock {\em Physical Review Letters\/} {\bf 113} (2014) 262504.
\newblock \doi{10.1103/PhysRevLett.113.262504}.

\bibitem[{Sargsyan et~al.(2026)Sargsyan, King, Glick-Magid, and Seng}]{Sargsyan2026}
Sargsyan GH, King GB, Glick-Magid A, Seng CY.
\newblock The role of ab initio beta-decay calculations in light nuclei for probes of physics beyond the standard model  (2026).

\bibitem[{Gysbers et~al.(2019)Gysbers, Hagen, Holt, Jansen, Morris, Navr{\'{a}}til et~al.}]{Gysbers2019}
Gysbers P, Hagen G, Holt JD, Jansen GR, Morris TD, Navr{\'{a}}til P, et~al.
\newblock {Discrepancy between experimental and theoretical $\beta$-decay rates resolved from first principles}.
\newblock {\em Nature Physics\/} {\bf 15} (2019) 428--431.
\newblock \doi{10.1038/s41567-019-0450-7}.

\bibitem[{Acharya et~al.(2025)Acharya, Sobczyk, Bacca, Hagen, and Jiang}]{Acharya2025}
Acharya B, Sobczyk JE, Bacca S, Hagen G, Jiang W.
\newblock O$^{16}$ electroweak response functions from first principles.
\newblock {\em Physical Review Letters\/} {\bf 134} (2025) 202501.
\newblock \doi{10.1103/PhysRevLett.134.202501}.

\bibitem[{Hammer et~al.(2020)Hammer, König, and van Kolck}]{Hammer2019}
Hammer HW, König S, van Kolck U.
\newblock Nuclear effective field theory: Status and perspectives.
\newblock {\em Reviews of Modern Physics\/} {\bf 92} (2020) 025004.
\newblock \doi{10.1103/RevModPhys.92.025004}.

\bibitem[{Epelbaum et~al.(2020)Epelbaum, Krebs, and Reinert}]{Epelbaum2020}
Epelbaum E, Krebs H, Reinert P.
\newblock {High-Precision Nuclear Forces From Chiral EFT: State-of-the-Art, Challenges, and Outlook}.
\newblock {\em Frontiers in Physics\/} {\bf 8} (2020) 1--30.
\newblock \doi{10.3389/fphy.2020.00098}.

\bibitem[{Behr et~al.(2005)Behr, Gorelov, Melconian, Trinczek, Alford, Ashery et~al.}]{Behr2005}
Behr Ja, Gorelov a, Melconian D, Trinczek M, Alford WP, Ashery D, et~al.
\newblock {Weak interaction symmetries with atom traps}.
\newblock {\em European Physical Journal A\/} {\bf 25} (2005) 685--689.
\newblock \doi{10.1140/epjad/i2005-06-097-9}.

\bibitem[{Behr and Gwinner(2009)}]{Behr2008}
Behr JA, Gwinner G.
\newblock Standard model tests with trapped radioactive atoms.
\newblock {\em Journal of Physics G: Nuclear and Particle Physics\/} {\bf 36} (2009) 033101.
\newblock \doi{10.1088/0954-3899/36/3/033101}.

\bibitem[{Oblath and VanDevender(2024)}]{Oblath2024}
Oblath NS, VanDevender BA.
\newblock Cyclotron radiation emission spectroscopy.
\newblock {\em Annual Review of Nuclear and Particle Science\/} {\bf 74} (2024) 447--472.
\newblock \doi{10.1146/annurev-nucl-120523-021323}.

\bibitem[{Byron et~al.(2023)Byron, Harrington, Taylor, Degraw, Buzinsky, Dodson et~al.}]{Byron2023}
Byron W, Harrington H, Taylor RJ, Degraw W, Buzinsky N, Dodson B, et~al.
\newblock {First Observation of Cyclotron Radiation from MeV-Scale e± following Nuclear $\beta$ Decay}.
\newblock {\em Physical Review Letters\/} {\bf 131} (2023).
\newblock \doi{10.1103/PhysRevLett.131.082502}.

\bibitem[{Carney et~al.(2023)Carney, Leach, and Moore}]{Carney2023}
Carney D, Leach KG, Moore DC.
\newblock Searches for massive neutrinos with mechanical quantum sensors.
\newblock {\em PRX Quantum\/} {\bf 4} (2023) 010315.
\newblock \doi{10.1103/PRXQuantum.4.010315}.

\bibitem[{Smolsky et~al.(2025)Smolsky, Leach, Abells, Amaro, Andoche, Borbridge et~al.}]{Smolsky2024}
Smolsky J, Leach KG, Abells R, Amaro P, Andoche A, Borbridge K, et~al.
\newblock Direct experimental constraints on the spatial extent of a neutrino wavepacket.
\newblock {\em Nature\/} {\bf 638} (2025) 640--644.
\newblock \doi{10.1038/s41586-024-08479-6}.

\bibitem[{Friedrich et~al.(2021)Friedrich, Kim, Bray, Cantor, Dilling, Fretwell et~al.}]{Friedrich2021}
Friedrich S, Kim GB, Bray C, Cantor R, Dilling J, Fretwell S, et~al.
\newblock Limits on the existence of sub-mev sterile neutrinos from the decay of $^7$be in superconducting quantum sensors.
\newblock {\em Physical Review Letters\/} {\bf 126} (2021) 021803.
\newblock \doi{10.1103/PhysRevLett.126.021803}.

\bibitem[{Hayen et~al.(2018)Hayen, Severijns, Bodek, Rozpedzik, and Mougeot}]{Hayen2018}
Hayen L, Severijns N, Bodek K, Rozpedzik D, Mougeot X.
\newblock {High precision analytical description of the allowed beta spectrum shape}.
\newblock {\em Reviews of Modern Physics\/} {\bf 90} (2018) 015008.
\newblock \doi{10.1103/RevModPhys.90.015008}.

\bibitem[{Hayen and Severijns(2019{\natexlab{a}})}]{Hayen2019a}
Hayen L, Severijns N.
\newblock {Beta Spectrum Generator: High precision allowed $\beta$ spectrum shapes}.
\newblock {\em Computer Physics Communications\/} {\bf 240} (2019{\natexlab{a}}) 152--164.
\newblock \doi{10.1016/j.cpc.2019.02.012}.

\bibitem[{Hayen(2024)}]{Hayen2024}
Hayen L.
\newblock Opportunities and open questions in modern beta decay.
\newblock {\em Annual Reviews of Nuclear and Particle Science\/} {\bf 09} (2024) 497--528.
\newblock \doi{10.1146/annurev-nucl-121423}.

\bibitem[{Hayen et~al.(2019{\natexlab{a}})Hayen, Kostensalo, Severijns, and Suhonen}]{Hayen2019}
Hayen L, Kostensalo J, Severijns N, Suhonen J.
\newblock {First-forbidden transitions in reactor antineutrino spectra}.
\newblock {\em Physical Review C\/} {\bf 99} (2019{\natexlab{a}}) 031301(R).
\newblock \doi{10.1103/PhysRevC.99.031301}.

\bibitem[{Hayen et~al.(2019{\natexlab{b}})Hayen, Kostensalo, Severijns, and Suhonen}]{Hayen2019b}
Hayen L, Kostensalo J, Severijns N, Suhonen J.
\newblock {First-forbidden transitions in the reactor anomaly}.
\newblock {\em Physical Review C\/} {\bf 100} (2019{\natexlab{b}}) 054323.
\newblock \doi{10.1103/PhysRevC.100.054323}.

\bibitem[{Zhang et~al.(2025)Zhang, Irani, Mendenhall, Rybicki, Hayen, Bowden et~al.}]{Zhang2025}
Zhang X, Irani A, Mendenhall MP, Rybicki N, Hayen L, Bowden N, et~al.
\newblock Conflux: A standardized framework to calculate reactor antineutrino flux.
\newblock {\em Computer Physics Communications\/} {\bf 317} (2025) 109831.
\newblock \doi{10.1016/j.cpc.2025.109831}.

\bibitem[{Hayen(2025)}]{Hayen2025}
Hayen L.
\newblock Impact of reactor neutrino uncertainties on coherent scattering’s discovery potential.
\newblock {\em Journal of Physics G: Nuclear and Particle Physics\/} {\bf 52} (2025) 055103.
\newblock \doi{10.1088/1361-6471/ad8ee2}.

\bibitem[{Konopinski and Uhlenbeck(1935)}]{Konopinski1935}
Konopinski EJ, Uhlenbeck GE.
\newblock {On the Fermi Theory of $\beta$-Radioactivity}.
\newblock {\em Physical Review\/} {\bf 48} (1935) 7.

\bibitem[{Konopinski and Uhlenbeck(1941)}]{Konopinski1941}
Konopinski EJ, Uhlenbeck GE.
\newblock {On the Fermi Theory of $\beta$-Radioactivity. II. The "Forbidden" Spectra}.
\newblock {\em Physical Review\/} {\bf 60} (1941) 308.

\bibitem[{Konopinski(1943)}]{Konopinski1943}
Konopinski EJ.
\newblock {Beta- Decay}.
\newblock {\em Reviews of Modern Physics\/} {\bf 15} (1943) 209.

\bibitem[{Holstein(1974{\natexlab{b}})}]{Holstein1974a}
Holstein BR.
\newblock {Electromagnetic corrections to allowed nuclear beta decay}.
\newblock {\em Physical Review C\/} {\bf 9} (1974{\natexlab{b}}) 1742.
\newblock \doi{10.1103/PhysRevC.9.1742}.

\bibitem[{Holstein(1974{\natexlab{c}})}]{Holstein1974b}
Holstein BR.
\newblock {Induced Coulomb corrections to nuclear beta decay}.
\newblock {\em Physical Review C\/} {\bf 10} (1974{\natexlab{c}}) 1215--1219.
\newblock \doi{10.1103/PhysRevC.10.1215}.

\bibitem[{Calaprice and Holstein(1976)}]{Calaprice1976}
Calaprice FP, Holstein BR.
\newblock Weak magnetism and the beta spectra of $^{12}$b and $^{12}$n.
\newblock {\em Nuclear Physics A\/} {\bf 273} (1976) 301--325.

\bibitem[{Kleppinger et~al.(1977)Kleppinger, Calaprice, and Holstein}]{Kleppinger1977}
Kleppinger W, Calaprice F, Holstein BR.
\newblock {Analysis of the recoil ion spectra for the $\beta$-decays of 6He, 19Ne and 35Ar}.
\newblock {\em Nuclear Physics A\/} {\bf 293} (1977) 46--60.
\newblock \doi{10.1016/0375-9474(77)90475-4}.

\bibitem[{Gorelov et~al.(2005)Gorelov, Melconian, Alford, Ashery, Ball, Behr et~al.}]{Gorelov2005}
Gorelov a, Melconian D, Alford WP, Ashery D, Ball G, Behr Ja, et~al.
\newblock {Scalar interaction limits from the $\beta$-$\nu$ correlation of trapped radioactive atoms}.
\newblock {\em Phys. Rev. Lett.\/} {\bf 94} (2005) 3--6.
\newblock \doi{10.1103/PhysRevLett.94.142501}.

\bibitem[{Wauters et~al.(2010)Wauters, Kraev, Z{\'{a}}kouck{\'{y}}, Beck, Breitenfeldt, {De Leebeeck} et~al.}]{Wauters2010}
Wauters F, Kraev I, Z{\'{a}}kouck{\'{y}} D, Beck M, Breitenfeldt M, {De Leebeeck} V, et~al.
\newblock Precision measurements of the $^{60}$co $\beta$-asymmetry parameter in search for tensor currents in weak interactions.
\newblock {\em Physical Review C\/} {\bf 82} (2010) 055502.
\newblock \doi{10.1103/PhysRevC.82.055502}.

\bibitem[{Soti et~al.(2014)Soti, Wauters, Breitenfeldt, Finlay, Herzog, Knecht et~al.}]{Soti2014}
Soti G, Wauters F, Breitenfeldt M, Finlay P, Herzog P, Knecht A, et~al.
\newblock Measurement of the $\beta$-asymmetry parameter of $^{67}$cu in search for tensor-type currents in the weak interaction.
\newblock {\em Physical Review C - Nuclear Physics\/} {\bf 90} (2014) 035502.
\newblock \doi{10.1103/PhysRevC.90.035502}.

\bibitem[{Brown et~al.(2018)Brown, Dees, Adamek, Allgeier, Blatnik, Bowles et~al.}]{Brown2018}
Brown MA, Dees EB, Adamek E, Allgeier B, Blatnik M, Bowles TJ, et~al.
\newblock {New result for the neutron $\beta$ -asymmetry parameter A0 from UCNA}.
\newblock {\em Physical Review C\/} {\bf 97} (2018) 35505.
\newblock \doi{10.1103/PhysRevC.97.035505}.

\bibitem[{Burkey et~al.(2022)Burkey, Savard, Gallant, Scielzo, Clark, Hirsh et~al.}]{Burkey2022}
Burkey MT, Savard G, Gallant AT, Scielzo ND, Clark JA, Hirsh TY, et~al.
\newblock Improved limit on tensor currents in the weak interaction from li 8 $\beta$ decay.
\newblock {\em Physical Review Letters\/} {\bf 128} (2022) 202502.
\newblock \doi{10.1103/PhysRevLett.128.202502}.

\bibitem[{Severijns et~al.(2023)Severijns, Hayen, {De Leebeeck}, Vanlangendonck, Bodek, Rozpedzik et~al.}]{Severijns2023}
Severijns N, Hayen L, {De Leebeeck} V, Vanlangendonck S, Bodek K, Rozpedzik D, et~al.
\newblock Ft values of the $\beta$ mirror itransitions and the weak-magnetism-induced current in allowed nuclear $\beta$ decay.
\newblock {\em Physical Review C\/} {\bf 107} (2023) 015502.
\newblock \doi{10.1103/PhysRevC.107.015502}.

\bibitem[{Sargsyan et~al.(2022)Sargsyan, Launey, Burkey, Gallant, Scielzo, Savard et~al.}]{Sargsyan2022}
Sargsyan GH, Launey KD, Burkey MT, Gallant AT, Scielzo ND, Savard G, et~al.
\newblock {Impact of Clustering on the Li 8 $\beta$ Decay and Recoil Form Factors}.
\newblock {\em Physical Review Letters\/} {\bf 128} (2022) 202503.
\newblock \doi{10.1103/PhysRevLett.128.202503}.

\bibitem[{Durand et~al.(1962)Durand, DeCelles, and Marr}]{Durand1962}
Durand L, DeCelles PC, Marr RB.
\newblock {Lorentz Invariance and the Kinematic Structure of Vertex Functions}.
\newblock {\em Physical Review\/} {\bf 126} (1962) 1882--1898.
\newblock \doi{10.1103/PhysRev.126.1882}.

\bibitem[{Galindo and Pascual(1969)}]{Galindo1969}
Galindo A, Pascual P.
\newblock {On the elementary particle approach to muon capture}.
\newblock {\em Nuclear Physics, Section B\/} {\bf 14} (1969) 37--51.
\newblock \doi{10.1016/0550-3213(69)90342-3}.

\bibitem[{Delorme(1979)}]{Delorme1979}
Delorme J.
\newblock {\em Microscopic analysis of the elementary particle description of nuclear currents\/} (North Holland Publ. Co. Amsterdam) (1979).

\bibitem[{Jacob and Wick(1959)}]{Jacob1959}
Jacob M, Wick G.
\newblock On the general theory of collisions for particles with spin.
\newblock {\em Annals of Physics\/} {\bf 7} (1959) 404--428.
\newblock \doi{10.1016/0003-4916(59)90051-X}.

\bibitem[{de~Forest and Walecka(1966)}]{DeForest1966}
de~Forest T, Walecka JD.
\newblock {Electron scattering and nuclear structure}.
\newblock {\em Advances in Physics\/} {\bf 15} (1966) 1--109.
\newblock \doi{10.1080/00018736600101254}.

\bibitem[{Donnelly and Walecka(1975)}]{Donnelly1975}
Donnelly TW, Walecka JD.
\newblock {Electron Scattering and Nuclear Structure}.
\newblock {\em Annual Review of Nuclear Science\/} {\bf 25} (1975) 329--405.
\newblock \doi{10.1146/annurev.ns.25.120175.001553}.

\bibitem[{Walecka(2004)}]{Walecka2004}
Walecka JD.
\newblock {\em {Theoretical Nuclear and Subnuclear Physics}\/} (World Scientific), 2nd editio edn. (2004).

\bibitem[{King et~al.(2023)King, Baroni, Cirigliano, Gandolfi, Hayen, Mereghetti et~al.}]{King2022}
King GB, Baroni A, Cirigliano V, Gandolfi S, Hayen L, Mereghetti E, et~al.
\newblock Ab initio calculation of the $\beta$-decay spectrum of he6.
\newblock {\em Physical Review C\/} {\bf 107} (2023) 015503.
\newblock \doi{10.1103/PhysRevC.107.015503}.

\bibitem[{Glick-Magid et~al.(2022)Glick-Magid, Forss{\'{e}}n, Gazda, Gazit, Gysbers, and Navr{\'{a}}til}]{Glick-Magid2022}
Glick-Magid A, Forss{\'{e}}n C, Gazda D, Gazit D, Gysbers P, Navr{\'{a}}til P.
\newblock {Nuclear ab initio calculations of 6He $\beta$-decay for beyond the Standard Model studies}.
\newblock {\em Physics Letters B\/} {\bf 832} (2022) 137259.
\newblock \doi{10.1016/j.physletb.2022.137259}.

\bibitem[{Hardy and Towner(2020)}]{Hardy2020}
Hardy JC, Towner IS.
\newblock Superallowed $0^+ \to 0+$ nuclear $\beta$ decays: 2020 critical survey, with implications for $v_{ud}$ and ckm unitarity.
\newblock {\em Physical Review C\/} {\bf 102} (2020) 045501.
\newblock \doi{10.1103/PhysRevC.102.045501}.

\bibitem[{Edmonds and Mendlowitz(1957)}]{Edmonds1957}
Edmonds AR, Mendlowitz H.
\newblock {\em {Angular Momentum in Quantum Mechanics}\/} (Princeton, New Jersey: Princeton University) (1957).

\bibitem[{Hayen and Young(2020)}]{Hayen2020a}
Hayen L, Young AR.
\newblock {Consistent description of angular correlations in $\beta$ decay for Beyond Standard Model physics searches}  (2020).

\bibitem[{Behrens et~al.(1978)Behrens, Genz, Conze, Feldmeier, Stock, and Richter}]{Behrens1978}
Behrens H, Genz H, Conze M, Feldmeier H, Stock W, Richter A.
\newblock {Allowed $\beta$-transitions, weak magnetism and nuclear structure in light nuclei}.
\newblock {\em Annals of Physics\/} {\bf 115} (1978) 276--324.
\newblock \doi{10.1016/0003-4916(78)90158-6}.

\bibitem[{Rose and Osborn(1954{\natexlab{b}})}]{Rose1954a}
Rose ME, Osborn RK.
\newblock {The Pseudoscalar Interaction and the Beta Spectrum of RaE}.
\newblock {\em Phys. Rev.\/} {\bf 93} (1954{\natexlab{b}}) 1315--1325.

\bibitem[{Foldy and Wouthuysen(1950)}]{Foldy1950}
Foldy LL, Wouthuysen SA.
\newblock {On the Dirac Theory of Spin 1/2 Particles and Its Non-Relativistic Limit}.
\newblock {\em Physical Review\/} {\bf 78} (1950) 29--36.
\newblock \doi{10.1103/PhysRev.78.29}.

\bibitem[{Wilkinson(1982)}]{Wilkinson1982}
Wilkinson D.
\newblock {Analysis of neutron $\beta$-decay}.
\newblock {\em Nuclear Physics A\/} {\bf 377} (1982) 474--504.
\newblock \doi{10.1016/0375-9474(82)90051-3}.

\bibitem[{Shekhter(1959)}]{Shekhter1959}
Shekhter VM.
\newblock {Hyperon Beta Decay}.
\newblock {\em Soviet Physics JETP\/} {\bf 35} (1959) 316.

\bibitem[{Glick-Magid et~al.(2016)Glick-Magid, Mishnayot, Mukul, Hass, Ron, Vaintraub et~al.}]{Glick-Magid2016a}
Glick-Magid A, Mishnayot Y, Mukul I, Hass M, Ron G, Vaintraub S, et~al.
\newblock {Beta spectrum of unique first-forbidden decays as a novel test for fundamental symmetries}.
\newblock {\em Physics Letters B\/} {\bf 767} (2016) 285.
\newblock \doi{10.1016/j.physletb.2017.02.023}.

\bibitem[{Glick-Magid and Gazit(2023)}]{Glick-Magid2023}
Glick-Magid A, Gazit D.
\newblock {Multipole decomposition of tensor interactions of fermionic probes with composite particles and BSM signatures in nuclear reactions}.
\newblock {\em Physical Review D\/} {\bf 107} (2023) 075031.
\newblock \doi{10.1103/PhysRevD.107.075031}.

\bibitem[{Seng et~al.(2025)Seng, Glick-Magid, and Cirigliano}]{Seng2025}
Seng CY, Glick-Magid A, Cirigliano V.
\newblock Unique forbidden beta decays at zero momentum transfer.
\newblock {\em Physical Review Letters\/} {\bf 134} (2025) 081805.
\newblock \doi{10.1103/PhysRevLett.134.081805}.

\bibitem[{Severijns et~al.(2008)Severijns, Tandecki, Phalet, and Towner}]{Severijns2008}
Severijns N, Tandecki M, Phalet T, Towner IS.
\newblock $\mathcal{F}t$ values of the $t=1/2$ mirror $\beta$ transitions.
\newblock {\em Physical Review C\/} {\bf 78} (2008) 055501.
\newblock \doi{10.1103/PhysRevC.78.055501}.

\bibitem[{Naviliat-Cuncic and Severijns(2009)}]{Naviliat-Cuncic2009}
Naviliat-Cuncic O, Severijns N.
\newblock {Test of the Conserved Vector Current Hypothesis in $T=1/2$ Mirror Transitions and New Determination of $|V{\_}{\{}ud{\}}|$}.
\newblock {\em Physical Review Letters\/} {\bf 102} (2009) 142302.
\newblock \doi{10.1103/PhysRevLett.102.142302}.

\bibitem[{Fenker et~al.(2018)Fenker, Gorelov, Melconian, Behr, Anholm, Ashery et~al.}]{Fenker2018}
Fenker B, Gorelov A, Melconian D, Behr JA, Anholm M, Ashery D, et~al.
\newblock Precision measurement of the $\beta$ asymmetry in spin-polarized $^{37}$k decay.
\newblock {\em Physical Review Letters\/} {\bf 120} (2018) 062502.
\newblock \doi{10.1103/PhysRevLett.120.062502}.

\bibitem[{Sternberg et~al.(2015)Sternberg, Segel, Scielzo, Savard, Clark, Bertone et~al.}]{Sternberg2015}
Sternberg MGG, Segel R, Scielzo NDD, Savard G, Clark JAA, Bertone PFF, et~al.
\newblock Limit on tensor currents from $^8$li $\beta$ decay.
\newblock {\em Physical Review Letters\/} {\bf 115} (2015) 182501.
\newblock \doi{10.1103/PhysRevLett.115.182501}.

\bibitem[{Gallant et~al.(2023)Gallant, Scielzo, Savard, Clark, Brodeur, Buchinger et~al.}]{Gallant2023}
Gallant AT, Scielzo ND, Savard G, Clark JA, Brodeur M, Buchinger F, et~al.
\newblock {Angular Correlations in the $\beta$ Decay of B 8: First Tensor-Current Limits from a Mirror-Nucleus Pair}.
\newblock {\em Physical Review Letters\/} {\bf 130} (2023) 192502.
\newblock \doi{10.1103/PhysRevLett.130.192502}.

\bibitem[{Triambak et~al.(2017)Triambak, Phuthu, Garc{\'{i}}a, Harper, Orce, Short et~al.}]{Triambak2017}
Triambak S, Phuthu L, Garc{\'{i}}a A, Harper GC, Orce JN, Short DA, et~al.
\newblock The $2_1^+ \to 3^+_1$ gamma width in $^{22}$na and second class currents.
\newblock {\em Physical Review C\/} {\bf 95} (2017) 035501.
\newblock \doi{10.1103/PhysRevC.95.035501}.

\bibitem[{Hardy and Towner(2005)}]{Hardy2004}
Hardy JC, Towner IS.
\newblock Superallowed $0^+\to 0^+$ nuclear $\beta$ decays: A critical review.
\newblock {\em Physical Review C\/} {\bf 71} (2005) 055501.
\newblock \doi{10.1103/PhysRevC.71.055501}.

\bibitem[{Holstein(1979{\natexlab{a}})}]{Holstein1979}
Holstein BR.
\newblock {Electromagnetic effects and weak form factors}.
\newblock {\em Physical Review C\/} {\bf 19} (1979{\natexlab{a}}) 1467--1472.
\newblock \doi{10.1103/PhysRevC.19.1467}.

\bibitem[{Holstein(1979{\natexlab{b}})}]{Holstein1979b}
Holstein BR.
\newblock {Comment on Coulomb corrections to weak transition amplitudes}.
\newblock {\em Physical Review C\/} {\bf 19} (1979{\natexlab{b}}) 1544--1546.
\newblock \doi{10.1103/PhysRevC.19.1544}.

\bibitem[{Hayen(2021)}]{Hayen2021}
Hayen L.
\newblock Standard model $\mathcal{O}(\alpha)$ renormalization of $g_a$ and its impact on new physics searches.
\newblock {\em Physical Review D\/} {\bf 103} (2021) 113001.
\newblock \doi{10.1103/PhysRevD.103.113001}.

\bibitem[{Gorchtein(2019)}]{Gorchtein2018}
Gorchtein M.
\newblock {{\textless}math display="inline"{\textgreater} {\textless}mrow{\textgreater} {\textless}mi{\textgreater}$\gamma${\textless}/mi{\textgreater} {\textless}mi{\textgreater}W{\textless}/mi{\textgreater} {\textless}/mrow{\textgreater} {\textless}/math{\textgreater} Box Inside Out: Nuclear Polarizabilities Distort the Beta Decay Spectrum}.
\newblock {\em Physical Review Letters\/} {\bf 123} (2019) 042503.
\newblock \doi{10.1103/PhysRevLett.123.042503}.

\bibitem[{Seng et~al.(2020)Seng, Galviz, and Mei{\ss}ner}]{Seng2019}
Seng CY, Galviz D, Mei{\ss}ner UG.
\newblock {A new theory framework for the electroweak radiative corrections in Kl3 decays}.
\newblock {\em Journal of High Energy Physics\/} {\bf 2020} (2020) 69.
\newblock \doi{10.1007/JHEP02(2020)069}.

\bibitem[{Seng(2021)}]{Seng2021c}
Seng CY.
\newblock {Radiative Corrections to Semileptonic Beta Decays: Progress and Challenges}.
\newblock {\em Particles\/} {\bf 4} (2021) 397--468.
\newblock \doi{10.3390/particles4040034}.

\bibitem[{Seng and Gorchtein(2023)}]{Seng2023}
Seng CY, Gorchtein M.
\newblock {Dispersive formalism for the nuclear structure correction $\delta$nS to the $\beta$ decay rate}.
\newblock {\em Physical Review C\/} {\bf 107} (2023) 1--22.
\newblock \doi{10.1103/PhysRevC.107.035503}.

\bibitem[{Bottino and Ciocchetti(1973)}]{Bottino1973b}
Bottino A, Ciocchetti G.
\newblock {Coulomb corrections to nuclear beta decay}.
\newblock {\em Physics Letters B\/} {\bf 43} (1973) 170--174.
\newblock \doi{10.1016/0370-2693(73)90261-X}.

\bibitem[{Bottino et~al.(1974)Bottino, Ciocchetti, and Kim}]{Bottino1974}
Bottino A, Ciocchetti G, Kim CW.
\newblock {Coulomb corrections to nuclear beta decay through induced terms}.
\newblock {\em Physical Review C\/} {\bf 9} (1974) 2052--2055.
\newblock \doi{10.1103/PhysRevC.9.2052}.

\bibitem[{Hayen and Severijns(2019{\natexlab{b}})}]{Hayen2019c}
Hayen L, Severijns N.
\newblock {Radiative corrections to Gamow-Teller decays}.
\newblock {\em arXiv\/}  (2019{\natexlab{b}}).

\bibitem[{Gorchtein and Seng(2021)}]{Gorchtein2021}
Gorchtein M, Seng CY.
\newblock {Dispersion relation analysis of the radiative corrections to gA in the neutron $\beta$-decay}.
\newblock {\em Journal of High Energy Physics\/} {\bf 2021} (2021) 53.
\newblock \doi{10.1007/JHEP10(2021)053}.

\bibitem[{Seng et~al.(2018)Seng, Gorchtein, Patel, and Ramsey-Musolf}]{Seng2018}
Seng CY, Gorchtein M, Patel HH, Ramsey-Musolf MJ.
\newblock {Reduced Hadronic Uncertainty in the Determination of $V{\_}{\{}ud{\}}$}.
\newblock {\em Physical Review Letters\/} {\bf 121} (2018) 241804.
\newblock \doi{10.1103/PhysRevLett.121.241804}.

\bibitem[{Gorchtein and Seng(2023)}]{Gorchtein2023}
Gorchtein M, Seng CY.
\newblock The standard model theory of neutron beta decay  (2023).
\newblock \doi{10.3390/universe9090422}.

\bibitem[{Seng et~al.(2019)Seng, Gorchtein, and Ramsey-Musolf}]{Seng2019b}
Seng CY, Gorchtein M, Ramsey-Musolf MJ.
\newblock {Dispersive evaluation of the inner radiative correction in neutron and nuclear $\beta$ decay}.
\newblock {\em Physical Review D\/} {\bf 100} (2019) 013001.
\newblock \doi{10.1103/PhysRevD.100.013001}.

\bibitem[{Towner(1992)}]{Towner1992a}
Towner I.
\newblock {The nuclear-structure dependence of radiative corrections in superallowed Fermi beta-decay}.
\newblock {\em Nuclear Physics A\/} {\bf 540} (1992) 478--500.
\newblock \doi{10.1016/0375-9474(92)90170-O}.

\bibitem[{Towner(1994)}]{Towner1994}
Towner I.
\newblock {Quenching of spin operators in the calculation of radiative corrections for nuclear beta decay}.
\newblock {\em Physics Letters B\/} {\bf 333} (1994) 13--16.
\newblock \doi{10.1016/0370-2693(94)91000-6}.

\bibitem[{Towner and Hardy(2002)}]{Towner2002}
Towner IS, Hardy JC.
\newblock {Calculated corrections to superallowed Fermi beta decay: New evaluation of the nuclear-structure-dependent terms}.
\newblock {\em Phys. Rev. C\/} {\bf 66} (2002) 35501.
\newblock \doi{10.1103/PhysRevC.66.035501}.

\bibitem[{Gennari et~al.(2025)Gennari, Drissi, Gorchtein, Navrátil, and Seng}]{Gennari2025}
Gennari M, Drissi M, Gorchtein M, Navrátil P, Seng CY.
\newblock Ab initio strategy for taming the nuclear-structure dependence of vud extractions: The c 10 → b 10 superallowed transition.
\newblock {\em Physical Review Letters\/} {\bf 134} (2025).
\newblock \doi{10.1103/PhysRevLett.134.012501}.

\bibitem[{Gorchtein and Seng(2024)}]{Gorchtein2024}
Gorchtein M, Seng CY.
\newblock Superallowed nuclear beta decays and precision tests of the standard model.
\newblock {\em Annual Review of Nuclear and Particle Science\/} {\bf 74} (2024) 23--47.
\newblock \doi{10.1146/annurev-nucl-102622-020726}.

\bibitem[{Lovato et~al.(2023)Lovato, Nikolakopoulos, Rocco, and Steinberg}]{Lovato2023}
Lovato A, Nikolakopoulos A, Rocco N, Steinberg N.
\newblock Lepton–nucleus interactions within microscopic approaches.
\newblock {\em Universe\/} {\bf 9} (2023) 367.
\newblock \doi{10.3390/universe9080367}.

\bibitem[{Rocco et~al.(2019)Rocco, Nakamura, Lee, and Lovato}]{Rocco2019}
Rocco N, Nakamura SX, Lee TSH, Lovato A.
\newblock Electroweak pion production on nuclei within the extended factorization scheme.
\newblock {\em Physical Review C\/} {\bf 100} (2019) 045503.
\newblock \doi{10.1103/PhysRevC.100.045503}.

\bibitem[{Rocco(2020)}]{Rocco2020}
Rocco N.
\newblock Ab initio calculations of lepton-nucleus scattering.
\newblock {\em Frontiers in Physics\/} {\bf 8} (2020).
\newblock \doi{10.3389/fphy.2020.00116}.

\bibitem[{Sobczyk(2024)}]{Sobczyk2024}
Sobczyk JE.
\newblock Electron scattering on $^4$he from coupled-cluster theory.
\newblock {\em Few-Body Systems\/} {\bf 65} (2024) 48.
\newblock \doi{10.1007/s00601-024-01913-5}.

\bibitem[{King and Pastore(2024)}]{King2024}
King GB, Pastore S.
\newblock Recent progress in the electroweak structure of light nuclei using quantum monte carlo methods.
\newblock {\em Annual Review of Nuclear and Particle Science\/} {\bf 74} (2024) 343--368.
\newblock \doi{10.1146/annurev-nucl-101920-021401}.

\end{thebibliography}


\end{document}